\documentclass[lettersize,journal]{IEEEtran}
\usepackage[utf8]{inputenc}
\DeclareUnicodeCharacter{03B4}{$\delta$}
\usepackage{amsmath,amsfonts}
\usepackage{algorithmic}
\usepackage{algorithm}
\usepackage{array}
\usepackage[caption=false,font=normalsize,labelfont=sf,textfont=sf]{subfig}
\usepackage{textcomp}
\usepackage{stfloats}
\usepackage{url}
\usepackage{verbatim}
\usepackage{graphicx}
\usepackage{cite}
\usepackage[colorlinks, citecolor=blue]{hyperref}
\DeclareUnicodeCharacter{2009}{\,} 

\begin{document}

\title{Lifecycle Management of Optical Networks with Dynamic-Updating Digital Twin: A Hybrid Data-Driven and Physics-Informed Approach}

\author{Yuchen Song, Min Zhang, Yao Zhang, Yan Shi, Shikui Shen, Xiongyan Tang, Shanguo Huang, and Danshi Wang,~\IEEEmembership{Senior member,~IEEE}
\thanks{This work was supported by National Natural Science Foundation of China No. 62171053, Beijing Nova Program No. 20230484331 (Corresponding author: Danshi Wang).}

\thanks{Yuchen Song, Min Zhang, Yao Zhang, Shanguo Huang, and Danshi Wang are with the State Key Laboratory of Information Photonics and Optical Communications, Beijing University of Posts and Telecommunications, Beijing 100876, China (e-mail: songyc@bupt.edu.cn; mzhang@bupt.edu.cn; zhang-yao@bupt.edu.cn; shghuang@bupt.edu.cn; danshi wang@bupt.edu.cn).}

\thanks{Yan Shi, Shikui Shen, and Xiongyan Tang are with China Unicom Research Institute, Beijing, 100044, China. (e-mail: shiyan49@chinaunicom.cn; shensk@chinaunicom.cn; tangxy@chinaunicom.cn).}

}

\markboth{IEEE Journal on Selected Areas in Communications,~Vol.~x, No.~x, April~2024}%
{Shell \MakeLowercase{\textit{et al.}}: A Sample Article Using IEEEtran.cls for IEEE Journals}

\IEEEpubid{}

\maketitle

\begin{abstract}
Digital twin (DT) techniques have been proposed for the autonomous operation and lifecycle management of next-generation optical networks. To fully utilize potential capacity and accommodate dynamic services, the DT must dynamically update in sync with deployed optical networks throughout their lifecycle, ensuring low-margin operation. This paper proposes a dynamic-updating DT for the lifecycle management of optical networks, employing a hybrid approach that integrates data-driven and physics-informed techniques for fiber channel modeling. This integration ensures both rapid calculation speed and high physics consistency in optical performance prediction while enabling the dynamic updating of critical physical parameters for DT. The lifecycle management of optical networks, covering accurate performance prediction at the network deployment and dynamic updating during network operation, is demonstrated through simulation in a large-scale network. Up to 100 times speedup in prediction is observed compared to classical numerical methods. In addition, the fiber Raman gain strength, amplifier frequency-dependent gain profile, and connector loss between fiber and amplifier on C and L bands can be simultaneously updated. Moreover, the dynamic-updating DT is verified on a field-trial C+L-band transmission link, achieving a maximum accuracy improvement of 1.4 dB for performance estimation post-device replacement. Overall, the dynamic-updating DT holds promise for driving the next-generation optical networks towards lifecycle autonomous management.  
\end{abstract}

\begin{IEEEkeywords}
Optical networks, Digital twin, Lifecycle management, Field-trial demonstration, Hybrid data-driven and physics-informed deep learning.
\end{IEEEkeywords}

\section{Introduction}
\IEEEPARstart{T}{he} rise of applications in the Internet of Things and artificial intelligence is imposing more stringent capacity and latency requirements on the underlying optical networks. Traditionally, most deployed optical networks are designed with large margins to counter short-term performance fluctuations and long-term component aging \cite{RN155}. To meet the ever-increasing traffic demand, researchers and operators are seeking to reduce these margins and fully utilize the potential capacity of optical networks, which is expected to yield significant economic benefits \cite{mitra2019effect}. Moreover, in the near future, optical fiber communications are poised to fully leverage the entire C+L-band transmission bandwidth and use more powerful optical cross-connects to deal with more dynamic service patterns \cite{essiambre2010capacity, hoshida2022ultrawideband, stepanovsky2019comparative}. With these developments, the next-generation optical networks are expected to operate in the low-margin domain with an accurate depiction of optical networks throughout their lifecycle \cite{RN155}. However, current network management through conventional simulations requires large safety margins because these simulations cannot trace changes in networks and thus cannot provide an accurate network description, leading to significant capacity waste \cite{soumplis2017network}. These transformative shifts in optical networks render conventional static network simulations inadequate, necessitating the implementation of advanced techniques with dynamic-updating capabilities for the lifecycle management of optical networks, especially during the operational stage.

As the digital replicas of optical networks in physical world, the digital twin (DT) is expected to use accurate mirroring models, comprehensive monitoring information, and real-time data transfer mechanisms to depict characteristics and predict the activities of optical networks over their lifetime \cite{RN158,RN120}. A fundamental distinction between DT and simulation is that DT aims to replicate realistic physical entities, whereas simulation typically predicts the behaviors of hypothetical systems. Additionally, simulations often involve one-time manual data flow from the physical entity, lacking direct reflection of any changes in the physical counterpart \cite{RN122}. However, seamless automatic interaction between DTs and physical systems should exist, facilitating the flow of monitoring data from the physical system to the DT and control commands from the DT to the physical system \cite{RN120}. Furthermore, the DT should generate an accurate depiction of time-varying network states during its lifecycle, implying the necessity of establishing a mechanism to dynamically update the DT in accordance with the real-time data collected from physical networks. These features make DT superior to conventional network simulations for the lifecycle management of next-generation optical networks to operate in low-margin region. In this connection, DT empowers network operators to enhance decision-making, optimize performance, and fortify the resilience of optical networks in an ever-evolving maintenance and operation landscape.

Currently, DTs under different frameworks have been demonstrated in both simulated and experimental optical networks, where the prediction of quality of transmission (QoT) estimation and various controls based on the predicted QoT were showcased \cite{wang2021role, vilalta2023applying, RN2,RN61}. However, the dynamic updating ability of DT is often overlooked in simulations and experiments conducted in ideal environments, where comprehensive measurements can be conducted and unexpected disturbances are minimal \cite{RN156}. Meanwhile, as the development of DT towards implementation on field-deployed networks, the significance of dynamic updating techniques becomes apparent. In practical imperfect environments, most parameters deviate from the nominal values provided by point-of-manufacture handbooks, often with pronounced additional impairments such as connector loss \cite{RN157}. Unexpected human activities and component ageing incur problems of parameter shifting in the lifecycle of optical networks \cite{pesic2019impact}. To implement a digital twin (DT) in field-deployed optical networks, it is particularly crucial to employ a dynamic-updating DT that aligns with the evolving network states in the physical world. 

In this paper, we extend our previous conference paper in \cite{song2023physics} and propose the dynamic-updating DT for the lifecycle management of optical networks. Our contributions include: 1) We introduce the hybrid data-driven and physics-informed deep operator network for both multi-channel optical power evolution prediction and QoT estimation in DT. This hybrid learning approach combines the fast calculation speed of data-driven methods with the high reliability of physics-informed methods. 2) With the integration of both data and physics, the DT can refine and update critical physical parameters using only a few measured channel power profiles, which can be conveniently collected by widely deployed optical channel monitors (OCMs). 3) To effectively balance the effects of data-driven and physics-informed loss functions, we propose a three-step training workflow for forward modeling and inverse updating. For demonstration, we first showcase the lifecycle management of optical networks using the proposed dynamic-updating DT in simulations with the COST 239 network topology. This includes precise prediction of channel power and QoT during the network deployment stage, as well as the ability to dynamically update physical parameters in response to network changes during the operation stage. After fully demonstrating the dynamic-updating DT on the simulation mesh, we implemented it in field-trial C+L-band optical transmission links, where practical restrictions and uncertainties exist. In the non-ideal environments of live production networks, the dynamic-updating DT improves the accuracy of channel power prediction and QoT estimation by updating the frequency-dependent amplifier gain profile, fiber Raman gain strength, and connector loss between fiber and amplifier on C and L bands. The dynamic-updating DT is expected to drive next-generation optical networks towards lifecycle autonomous management. 

The rest of the paper is organized as follows. In Section II, we offer a concise review of related works on DT, parameter identification techniques, and physics-informed machine learning. Section III explores the motivations behind lifecycle management of optical networks, emphasizing why physical parameters are the critical network status information to be updated and why the fiber channel is the most essential component in the models of DT. Section IV presents the hybrid modeling approach for fiber channel modeling. Section V demonstrates the dynamic-updating DT for the lifecycle management of a simulation network. In Section VI, the application of dynamic-updating DT is extended to a field-deployed C+L-band transmission link for updating after device replacement. Finally, we draw conclusions in Section VII.

\section{Related work}
The concept of DT was first proposed and developed in the area of aerospace \cite{RN158}, where the DT was used to mirror the life of its flying twin of a vehicle or system in physical world. The framework of DT in optical networks was proposed in \cite{wang2021role} encompassing the fusion of multi-source heterogenous monitoring data and utilization of models on fault management, hardware configuration, and transmission simulation. In \cite{vilalta2023applying}, the DT was established in conjunction with the software-defined network (SDN) controller, and the lifecycle of DT, including its deployment, updating, operation, and removal, was showcased. Subsequently, the DT, which aims at QoT estimation, in the SDN-based network operative system (NOS) has been demonstrated in both simulations and experiments \cite{RN2,RN61}. In \cite{wang2024digital}, techniques related to the telemetry, modeling, and self-learning of optical network DT were reviewed and discussed. Measurements informed models for DT in optical networks were summarized in \cite{RN122}, where measurements utilizing digital coherent transceivers and hybrid physics- and data-driven techniques are highlighted. Beyond simulations and experiments, the ultimate goal of DT is to provide tangible benefits to field-deployed production optical networks. QoT estimations have been demonstrated on field-deployed networks, focusing on the impact of connector loss \cite{RN157} and the benefits of monitoring techniques \cite{mackay2022field}. However, implementing DT on field-deployed networks faces challenges due to imperfect environments \cite{RN156}. Conducting detailed measurements of physical parameters in such networks, where devices and components are deployed across a large geographic scale and disturbances are frequent, is impractical. Therefore, accurate updating of physical parameters remains a key challenge.

To obtain accurate physical parameters, identification and refinement techniques in optical networks have been proposed. The refinement of connector loss, amplifier gain profile, and fiber types were proposed in \cite{morette2023machine}, where iterative gradient descent updates are conducted to refine these parameters with the goal of minimizing the prediction error of output power and signal-to-noise ratio (SNR) relative to the measured ones. Specifically for connector loss, its distribution was estimated with dichotomous algorithm by evaluating the stimulated Raman scattering (SRS) strength \cite{RN108}. In \cite{RN109}, channel powers measured by OCM at the output of erbium-doped fiber amplifiers (EDFA) were refined by calibrating with the total output power of EDFA, and the EDFA gain profile was also identified. The system drifts over time were investigated in \cite{RN109}, showing the importance of updating established DT. Other methods accounting for uncertainties in physical layer in a probabilistic way has been discussed \cite{RN160}. Moreover, longitudinal parameter identification techniques based on digital signal processing (DSP) have also gained popularity \cite{RN161}. However, a method capable of identifying multiple key physical parameters has yet to be demonstrated within the framework of DT, embodying its concept of dynamic-updating.

\begin{figure*}[t]
\centering\includegraphics[width=1\textwidth]{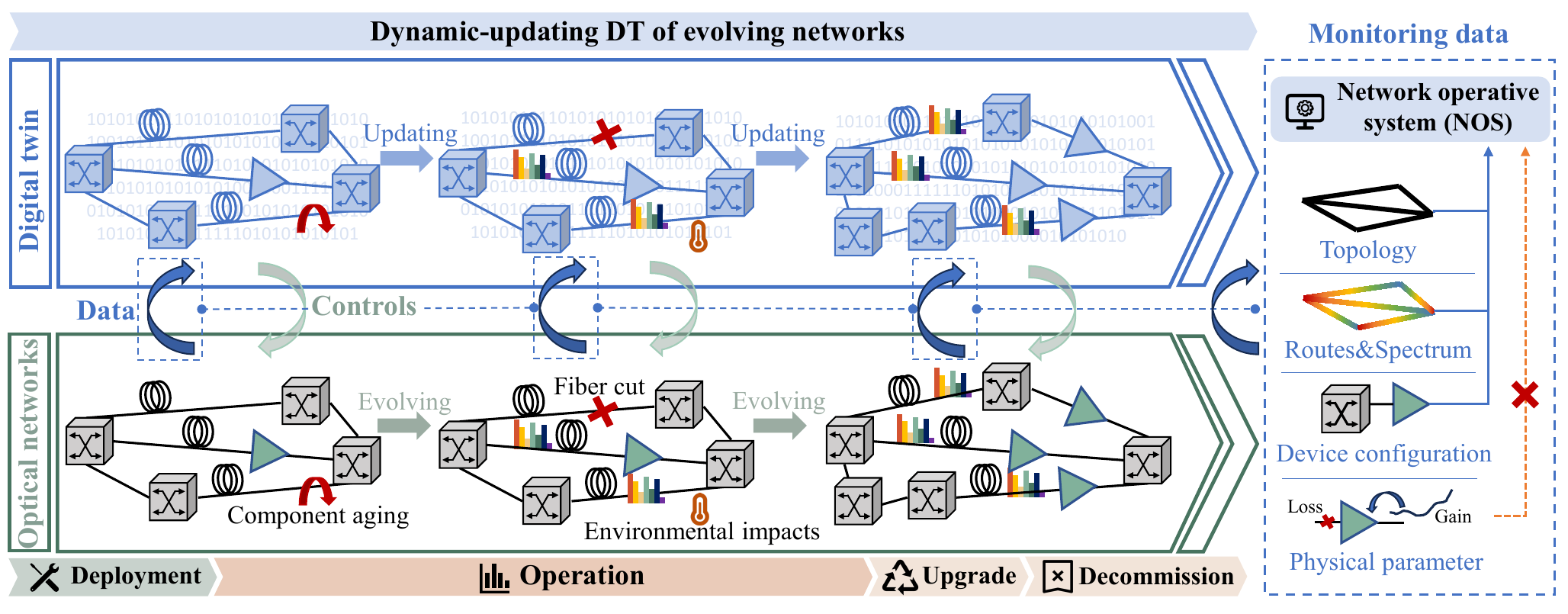}
\caption{Lifecycle management of optical networks using dynamic-updating DT with different types of monitoring data collected from optical networks.}
\label{fig1}
\end{figure*}

For models in the DT of optical networks, deep neural networks (NNs) have gain attentions due to their powerful ability in learning input-output mappings from data. Different from these data-driven approaches, physics-informed neural networks (PINNs) were proposed to directly solve partial differential equations (PDEs) without requiring any labeled data. The PDE is embedded in the loss function, and the derivatives of the PDE are calculated using automatic differentiation (AD) \cite{RN162}. PINN and its variants have been used in forward problems of modeling and predicting \cite{RN163, RN164, song2022physics} and inverse problems of identification or optimization \cite{lu2021physics,RN28,song2024srs}. The idea of physics-informed training methods has been recently introduced into NNs of new architecture, such as neural operator network, which greatly enhance the generalization ability. Furthermore, with practically available data of channel powers by OCMs, PINN can be used in refining parameters essential for constructing an accurate DT \cite{RN28}. Physics-informed machine learning (PIML) techniques offer a promising solution, as they enable the simultaneous conduction of forward prediction and inverse parameter updating using both data and physics. This makes them particularly suitable for dynamic-updating DT applications \cite{RN122}.

\section{Motivations}
The lifecycle of optical networks encompasses deployment, operation, and eventually upgrading to next-generation systems or decommissioning as legacy technology, as illustrated in Fig. 1. This entire process can span several years or even decades, with the operation stage occupying the majority of this timeframe. From the deployment stage onward, various practical challenges emerge in optical networks, underscoring the necessity for the dynamic-updating capability of DT. These challenges include environmental impacts, component aging, service changes, network failures, device upgrades, and more, all of which continually shape the evolving status of optical networks. For instance, service changes involving signal upload or download require configuration optimization of transceivers and amplifiers, and network failures due to fiber cuts necessitate protection switching to unused deployed fibers. During these stages, the DT should be continuously updated with the ever-evolving optical networks using the latest network status information, as illustrated in Fig. 1. Consequently, an accurate dynamic-updating DT can provide appropriate control commands for effective optical network management. In this paper, we primarily focus on the deployment and subsequent operation phases, as these are the most critical stages for network operators.

The monitoring data from the evolving network, transmitted to the DT, can be classified into different types, as shown in the right part of Fig. 1. Among these data, topology and device configuration can be automatically acquired by connecting nodes and devices online to the NOS \cite{vilalta2023applying}. Routing and spectrum resource information, on the other hand, can be captured by widely deployed sensors such as OCM and power monitor modules in transponders, with this data also being automatically routed to the NOS. However, data on physical parameters cannot be directly linked to the online NOS or identified by the various widely deployed monitors. Currently, information on fiber length and lump loss can be obtained using optical time domain reflectometers (OTDR). For other physical parameters, including characteristics of fiber, amplifier, and additional impairments such as connector loss, a cost-effective monitor that can be widely deployed has not yet been found. Therefore, accurate physical parameter identification poses significant challenges for updating the DT. At present, most physical parameters are updated manually with inaccurate nominal values recorded in the NOS.

To enable the updating of physical parameters, modeling of the fiber channel is an essential part of the DT. The channel power evolution $P_n$ in fiber on C+L-band transmission is described by a set of ordinary differential equations (ODEs), where the SRS is the prominent wideband nonlinear effect, as shown below: 

\begin{equation}
\begin{aligned}
    &\frac{\partial P_{n}(z)}{\partial z}+2 \alpha_{n} P_{n}(z)+ \\
    &r \sum_{m=1}^{N} \frac{g_{R}\left(f_{m}-f_{n}\right)}{A_{\text {eff }}} P_{n}(z) P_{m}(z)=0
\end{aligned}
\label{eq1}
\end{equation}

\begin{figure*}[t]
\centering\includegraphics[width=1\textwidth]{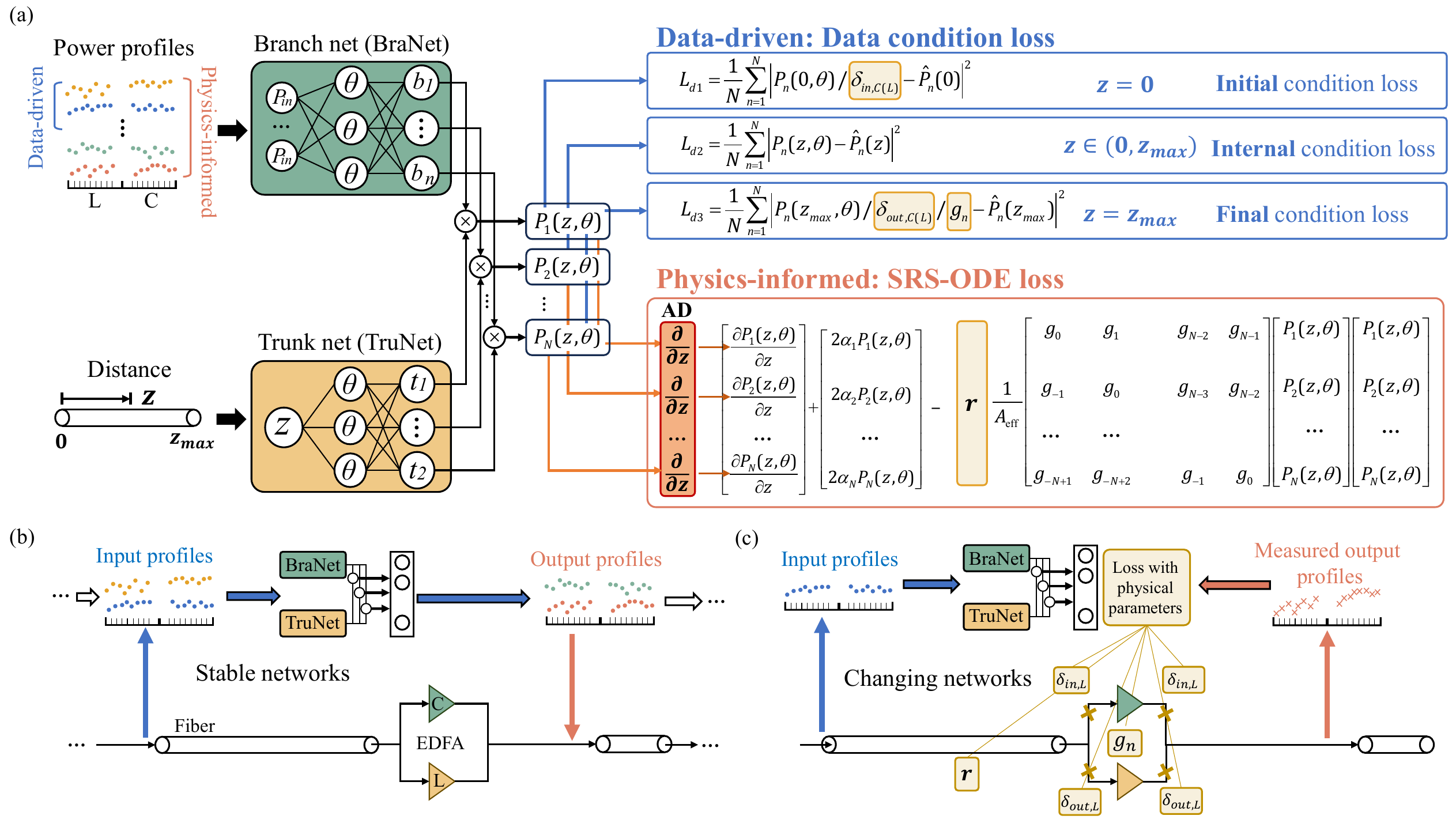}
\caption{Schematic of (a) hybrid data-driven and physics-informed DeepONet for fiber channel modeling, (b) trained DT for forward prediction in multi-span link, and (c) dynamic updating of the trained DT with changing networks (shifts on physical parameters) during lifecycle. Parameters to be updated are circled by yellow square.}
\label{fig3}
\end{figure*}

where $N$ is the number of transmitted channels, $\alpha_{n}$ is the frequency-dependent attenuation, $A_{eff}$ is the fiber effective area and $f_n$ denotes the frequency of the $n^{th}$ channel, $g_R$ is the fiber Raman gain spectrum, and $r$ is the strength of Raman gain that is directly related to the power transfer of SRS. Fiber channel entails the interaction among various linear and nonlinear effects. Specifically, forward power prediction, as described by Eq. (\ref{eq1}), can only be obtained through numerical iterative split-step methods, which is time-consuming, particularly in wideband large-scale optical networks with a large number of channels. The operational speed is critical for DT to enable performance prediction and optimization in time. On the other hand, for inverse parameter updating, the complex interplay of various effects within the fiber channel serves as the basis for identifying and refining multiple physical parameters along the entire link. For example, various techniques for parameter refinement were achieved by investigating the relationship between the input power of the fiber and the strength of Kerr and Raman effects within the fiber \cite{RN108,RN109}. Therefore, it is essential to employ appropriate learning techniques to understand the fiber channel in DT, enabling fast forward prediction and facilitating critical physical parameter updating throughout the entire link. In this context, PIML techniques emerge as promising solutions for fiber channel modeling, as depicted in Fig. \ref{fig3}. In PIML techniques, PDEs are integrated into the loss function of NNs, thereby reducing the need for extensive collection of data labels.

\section{Principles of dynamic-updating DT}
In the proposed dynamic-updating DT, the neural operator network is trained in a hybrid data-driven and physics-informed manner to learn the multi-channel power evolution of fiber channel. The established DT can not only improve the calculation speed of power evolution, but also remain physically plausible. In addition, the DT can be used for multiple critical physical parameters updating. 

\subsection{Physics-informed machine learning techniques and deep operator network}
In past years, NNs are rapidly gaining research interest in the field of optical networks. For most applications, NNs with sufficient representation ability are trained by massive paired input and output data, which is called the data-driven approach \cite{wang2020data}. Recently, with the advancement of PIML, PINN \cite{RN162} was proposed for obtaining solutions of PDEs by incorporating them as soft constraints in the loss function during the training. Thus, the PINN can be trained without using any labeled data to learn outputs satisfying the PDEs, which is called the physics-informed approach. Except for advancements in training approaches, new architectures have also been proposed. The Deep Operator Networks (DeepONet), developed based on the universal operator approximation theorem, differ from classical NNs based on the universal function approximation theorem \cite{lu2021learning}. As shown in Fig. \ref{fig3}, the structure of DeepONet comprises two NNs: the branch net (BraNet) and the trunk net (TruNet). The TruNet samples the transmission distance $z$ as inputs while the BraNet takes channel powers conditions of different loadings as input. The DeepONet outputs at given $z$, denoted as $P_n(z, \theta)$, are obtained by merging two net outputs by a vector product. The DeepONet was initially proposed to be trained in the data-driven framework \cite{lu2021learning}, meaning its outputs may not strictly adhere to physical laws and its training requires a large volume of data.

A purely data-driven approach or a purely physics-informed approach each has its own limitations. For the models in DT, the data-driven model can accurately describe the behaviors of actual optical networks by directly learning from sensor data. However, such data-driven models are only reliable within the region of input parameter space from which the data used to construct the model was taken. Using data-driven models for extrapolation without imposing any constraints based on physical knowledge can be dangerous. On the other hand, pure physics-informed models may have difficulty in fast calculation and has the possibility of following theoretical rules without following the actual situations in optical networks. The training of PINN is more computationally expensive compared to data-driven NNs because of the additional computations involved in calculating the PDE constraints. Integrating the physics-informed training approach to the architecture of DeepONet is expected to yield a new modeling technique characterized by fast calculation speed, high physical plausibility, and strong generalization ability \cite{wang2021learning}.

\subsection{Hybrid data-driven and physics-informed fiber channel}
We propose utilizing DeepONet to learn the channel power evolution operator (PEO) in the fiber channel through a hybrid data-driven and physics-informed manner. The PEO is a closed-form operator capable of mapping input power profiles to output profiles at required distances. As shown in the data-driven part of Fig. \ref{fig3}, the PEO can be straightforwardly trained through collected pairs of input and output channel power profiles with corresponding transmission distances. It should be highlighted that at distance 0 and $z_{max}$, the initial condition and final condition can be learned. The initial condition is necessary for power prediction, while the final condition is particularly important in afterwards parameter dynamic updating. At the same time, the outputs of PEO can be connected to the AD layer to calculate the derivatives of outputs with respect to the inputs. With additional physical parameter layer considering the $r$, the PEO can be guided by the SRS-ODE regularization as illustrated in Fig. \ref{fig3}, which is the physics-informed manner. For the hybrid training for forward prediction, all initial conditions are utilized for physics-informed learning with only a subset of these conditions paired with labeled outputs for data-driven learning. The SRS-ODE regularization $f$ of PEO is defined as follows:
\begin{equation}
\begin{aligned}
f_{n}(z, \theta, r)= \frac{\partial P_{n}(z, \theta)}{\partial z}+2 \alpha_{n} P_{n}(z, \theta)+ \\ r \sum_{m=1}^{N} \frac{g_{R}\left(f_{m}-f_{n}\right)}{A_{\text {eff }}} P_{n}(z, \theta) P_{m}(z, \theta)
\end{aligned}
\end{equation}

In this regularization, the channel power $P_n(z, \theta)$ predicted by PEO is not only dependent on the transmission distance $z$ but also dependent on the PEO parameters of weights and biases $\theta$. It should be noted that the SRS-ODE represents for a set of coupled equations with each one for a channel. Thus, the final SRS-ODE loss is calculated by a matrix as shown in Fig. \ref{fig3}. During the hybrid training, the PEO can be trained by minimising the mean squared error (MSE) of both data loss term $L_d$ and SRS-ODE loss term $L_f$ :
\begin{equation}
    L=\lambda_dL_d+\lambda_fL_f
\end{equation}
\begin{equation}
    L_d=\frac1{N_d}\sum_{i=1}^{N_d}\sum_{n=1}^N\lvert P_n(z_i,\theta)-\hat{P}_n(z_i)\rvert^2
\end{equation}
\begin{equation}
    L_{f}=\frac{1}{\boldsymbol{N}_{f}}\sum_{j=1}^{N_{f}}\sum_{n=1}^{N}\Bigl|f_{n}(z_{j},\theta,r)\Bigr|^{2}
\end{equation}

The data term $L_d$ represents the constraint imposed by measured data at distances denoted by $z_i$, while the SRS-ODE loss term $L_f$ enforces the regularization of $f_n(z, \theta, r)$ across the domain at random distances denoted by $z_j$. $\lambda_d$ and $\lambda_f$ are the corresponding two loss weights used to control the importance of the loss terms related to data and physics, respectively. Typically, the data term $L_d$ includes information of initial conditions and measurements at various distances in the forward prediction, or final conditions in the inverse updating, where the PEO outputs are trained to match the output labels $P_n$ at transmission distance $z_i$. On the other hand, the SRS-ODE loss $L_f$ integrates physical laws into loss functions and regularizes them across the entire transmission distance from input to output. Minimizing loss function $L$ by updating PEO parameter $\theta$ enables training in a hybrid data-driven and physics-informed manner. Notably, the Raman gain strength in $f_n(z, \theta, r)$ can also be trained in the inverse updating.

\begin{figure*}[t]
\centering\includegraphics[width=1\textwidth]{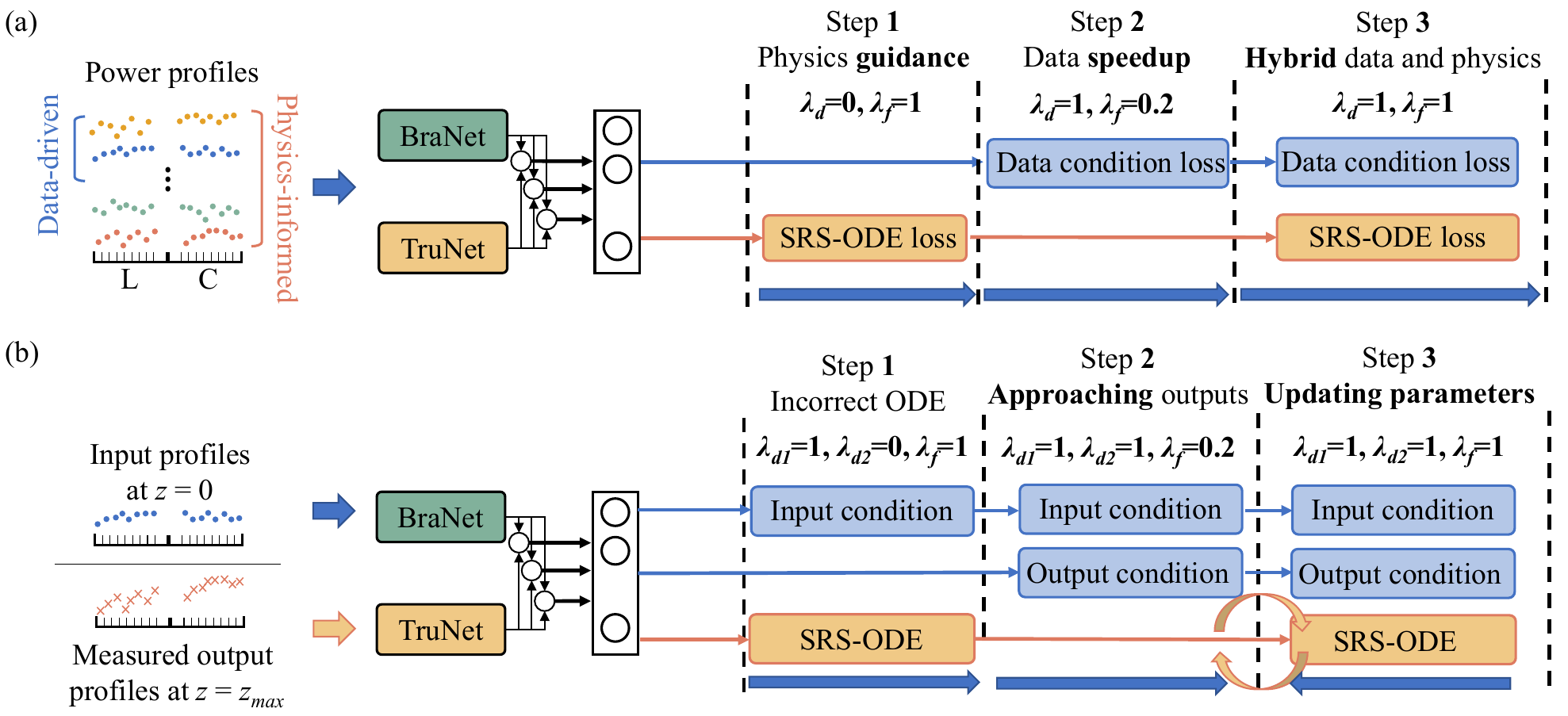}
\caption{Training workflow of using hybrid data-driven and physics-informed neural operator network in (a) forward channel power prediction, and (b) dynamic parameter updating.}
\label{fig4}
\end{figure*}

\subsection{Training workflow for forward power prediction}
In the forward power prediction, the PEO is trained as a closed-form neural operator in a hybrid data-driven and physics-informed manner. To learn the closed-form operator for multi-channel power evolution, initial conditions of input power profiles for training the PEO can be randomly generated. The generalization ability to unseen input power profiles can be significantly enhanced by the physics-informed learning. For the data-driven training part, we collect output labels corresponding to initial conditions at 20/40/60/80 and 100 km. These labels are used for minimizing the internal condition loss. In physics-informed learning, the SRS-ODE loss is regularized across the entire transmission distance.

To preserve the advantages of both data-driven and physics-informed training process, three distinct steps are involved, as depicted in Fig. \ref{fig4}(a). The first step involves guiding the PEO through physics-informed learning to a suitable point in the high-dimensional space of PEO parameters $\theta$, where the loss weight $\lambda_d$=0 and $\lambda_f$=1 in this step. Beginning training with pure physics-informed learning is crucial because initially, these two training approaches may not converge in the same direction. In the data-driven approach, the objective is to find a local optimal point that best maps the inputs to the outputs in the training dataset, with the outputs of the PEO forced to match the labels. However, the physics-informed approach aims to ensure that the PEO outputs satisfy the system of SRS-ODE governing the system's behavior. Simultaneously applying these two training approaches at the start of training can lead to unstable training. Moreover, the feature of physics-informed learning in ensuring that the PEO outputs satisfy the SRS-ODE instead of forcing them to match labels will greatly enhance the generalization ability \cite{wang2021learning}. Hence, we initially apply the physics-informed approach to establish a solid foundation. However, the feature of satisfying SRS-ODE also makes physics-informed learning slow in the training process. Thus, after the PEO has been guided to a suitable point (when the decrease of SRS-ODE loss levels off), data-driven learning is employed to expedite the training process, where the loss weight $\lambda_d$=1 and $\lambda_f$=0.2. A slight weighting on the SRS-ODE loss at step 2 ensures that the data-driven training does not steer the PEO to a completely different point in the high-dimensional parameter space of $\theta$. Finally, hybrid data-driven and physics-informed training is utilized to further refine and optimize the PEO, and the loss weight is set to be $\lambda_d$=1 and $\lambda_f$=1.

As illustrated in Fig. \ref{fig3}(b), the trained PEO, along with other models in the DT, can be utilized for the modeling of multi-span links. The input profiles of each span can be fed into the DT, and the corresponding outputs can be obtained after amplification, acting as the input for the next span.

\subsection{Training workflow for inverse parameter updating}
As depicted in Fig. \ref{fig3}(c), when optical networks undergo changes during lifecycle operation, such as shifts in certain physical parameters, the DT can be updated using the input and output power profiles of this span. In addition to updating the fiber Raman gain strength $r$, the connector losses at the fiber input and output for C(L)-band $\delta_{in,C(L)}$, $\delta_{out,C(L)}$, and the amplifier gain profile $g_n$ at channel $n$ along the transmission link can also be updated. To update this set of physical parameters $\Lambda$ = {$r$, $\delta_{in,C}$, $\delta_{in,L}$, $\delta_{out,C}$, $\delta_{out,L}$, $g_n$}, measured input and output power profiles, which can be easily collected by OCM, are required. In this case, for input and out power profiles at $z=0$ and $z_{max}$, the $L_d$ is as follows:
\begin{equation}
    L_d=\frac1N\sum_{n=1}^N\lvert P_n(0,\theta)/\delta_{in,C(L)}-\hat{P}_n(0)\rvert^2
\end{equation}
\begin{equation}
    L_d=\frac1N\sum_{n=1}^N\left|P_n(z_{max},\theta)/\delta_{out,C(L)}/g_n-\hat{P}_n(z_{max})\right|^2
\end{equation}

For channel $n$ in C-band, the $\delta_{in,C}$ is used; otherwise, $\delta_{in,L}$ should be used. In the inverse parameter updating, with measured input and output power profiles, physical parameters $\Lambda$ can be updated by minimizing the above data loss $L_d$ at $z=0$ and $z_{max}$ while satisfying the SRS-ODE loss. The whole loss function for inverse dynamic-updating is:
\begin{equation}
    L=\lambda_{d1}L_d(z=0)+\lambda_{d3}L_d(z=z_{max})+\lambda_fL_f
\end{equation}
where the loss weights $\lambda_{d1}$ and $\lambda_{d3}$ control the importance of initial and final condition loss, respectively. Physics-informed methods are well-suited for parameter updating tasks due to their inherent incorporation of physical parameters and ensuring the entire system satisfies the SRS-ODE. A pre-trained PEO is employed as it can provide a suitable starting point. $\Lambda$ are updated along with the network parameters $\theta$, ensuring the satisfaction of the SRS-ODE loss and the initial and final data condition loss at $z$=0 and $z=z_{max}$ as illustrated in Fig. \ref{fig3}.

Both the physical parameters $\Lambda$ and PEO parameters $\theta$ are updated in the inverse updating. To enhance the updating efficiency, three steps are involved in this process, as illustrated in Fig. \ref{fig4}(b). At start, physical parameters $\Lambda$ are incorrected. First, the initial condition loss and SRS-ODE loss with $\lambda_{d1}$=1, $\lambda_{d3}$=0, $\lambda_f$ =1 are utilized to ensure correctly learning of output power profiles with incorrect coefficients of SRS-ODE. In this step, the SRS-ODE loss is minimized by updating $\theta$ of PEO at $z$ ranging from 0 to $z_{max}$, and the $\Lambda$ do not update. In step 2, initial and final condition loss with $\lambda_{d1}$=1, $\lambda_{d3}$=1, $\lambda_f$=0.2 are used to reduce the difference between predicted power profiles and measured power profiles. At the step 3, all losses are used to update the $\Lambda$. Note that at the beginning of training the $\Lambda$ to be updated are incorrect. Therefore, in step 3, the SRS-ODE loss cannot be minimized properly with these incorrect parameters as long as the corresponding initial and final conditions are rigorously learned. This means that the simultaneous decrease in both the SRS-ODE loss and data loss contradicts the presence of incorrect $\Lambda$ at this step. After every 100 epochs in step 2 learning the initial and final conditions, the training is switched to step 3 for another 100 epochs to minimize the SRS-ODE loss by updating $\theta$, and the incorrect $\Lambda$ are also updating towards their actual values to minimize the SRS-ODE loss. In step 3, the loss weights are set to be $\lambda_{d1}$=1, $\lambda_{d3}$=1, $\lambda_f$=1. As both the PEO parameter $\theta$ and physical parameters $\Lambda$ need to be updated, to realize an efficient and stable training process, the physical parameters are mainly updated in step 3. The training iteratively alternates between step 2 and step 3, ensuring that both sets of parameters are refined and optimized over successive iterations. 

Based on forward prediction and inverse updating, the proposed dynamic-updating DT can be utilized for accurate performance evaluation of optical networks throughout their lifecycle. During greenfield deployment of optical networks, physical parameters can be refined, and the dynamic-updating DT can transition from a type-specialized DT, possessing nominal values of physical parameters, to an instance-specialized DT. This transition leads to more precise and tailored outcomes. In the afterwards brownfield maintenance, the dynamic-updating DT can alignment with shifted parameters of optical networks after unexpected disturbance. Different from the parameter identification techniques briefly reviewed in Section. II, the proposed dynamic-updating DT conduct forward prediction and inverse updating simultaneously, which is well-suited for the lifecycle management of next-generation optical networks. 

\begin{figure}[t]
\centering
\includegraphics[width=0.48\textwidth]{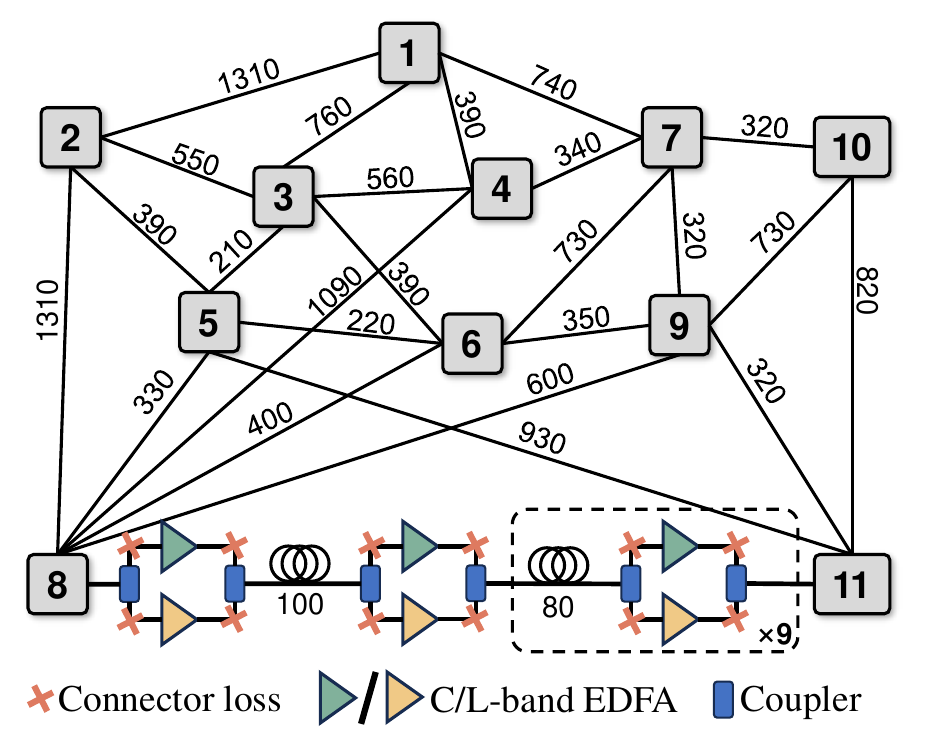}
\caption{Simulation networks with COST 239 topology. Coupler is used for coupling or splitting C and L band. Link between node 8 and 11 is shown as an example.}
\label{fig5}
\end{figure}

\section{Simulations on large-sacle optical networks}
We first demonstrate the proposed dynamic-updating DT for a part of the lifecycle management of optical networks through simulations. In the simulation network, we have the freedom to conduct any tests and obtain comprehensive results, illustrating the capability of the dynamic-updating DT. Additionally, we can test our proposed dynamic-updating DT on mesh networks and obtain statistical results. We select the COST 239 network topology, as depicted in Fig. \ref{fig5}. With a total of 11 nodes and 25 light paths between them, COST 239 is sufficiently large for evaluating the DT in terms of calculation speed and overall accuracy. The transmission bandwidth in this network is set to occupy the L-band, from 186.1 THz to 190.8 THz, and the C-band, from 191.4 THz to 196.1 THz, with a total of 96 channels.

\subsection{Models in DT}
QoT estimations are the primary outputs of the proposed DT. With the utilization of dual-polarization (DP) coherent optical technologies, each light path can be reliably approximated as an additive white Gaussian noise (AWGN) channel. In this regard, we adopt a decomposition-based approach for the models in DT, wherein the various impairments that collectively limit and determine transmission performance are assessed independently. In optical networks, the amplified spontaneous emission (ASE) noise from optical amplifiers, the nonlinear interference (NLI) from fiber propagation and the impairments from transceivers can be modeled as additive Gaussian disturbances at the receiver side \cite{buglia2022impact}. Therefore, the generalized signal-to-noise ratio (GSNR) can be defined as follows:

\begin{equation}
\label{eq9}
\begin{aligned}
& G S N R_{n}^{-1} \\
& \approx a^{-1}\left(S N R_{T R X}^{-1}+S N R_{A S E}^{-1}+S N R_{N L I}^{-1}\right) \\
& =a^{-1} \cdot\left(\frac{P_{n}}{\kappa_{n} P_{n}+P_{A S E, n}+P_{N L I, n}}\right)^{-1}
\end{aligned}
\end{equation}

where $SNR_{TRX}$, $SNR_{ASE}$, and $SNR_{NLI}$ are, respectively, the SNR from the transceiver subsystem or back-to-back implementation penalty, the ASE noise from the optical amplifier, and the accumulated NLI. $P_n$, $P_{ASE,n}$, and $P_{NLI,n}$ represent the channel power, ASE noise power, and NLI power of channel $n$, respectively. $\kappa_{i}=1 / S N R_{T R X}$ considers the effect of transceivers on GSNR. In addition to these noises, effects of the filtering penalty introduced by ROADMs can be summarized as a penalty coefficient on the final GSNR represented by a in Eq.(\ref{eq9}) \cite{buglia2022impact}. As one of the critical QoT metrics, the GSNR can be equivalently translated to pre-forward error correction (FEC) BER of optical networks in physical world, which is indispensable for margin calculation, device configuration, fault detection, upgrade planning \cite{borraccini2023experimental}, etc. To calculate this GSNR metric, the channel power evolution $P_n$ in fiber can be calculated by Eq.(\ref{eq1}).

\begin{figure}[t]
\centering
\includegraphics[width=0.48\textwidth]{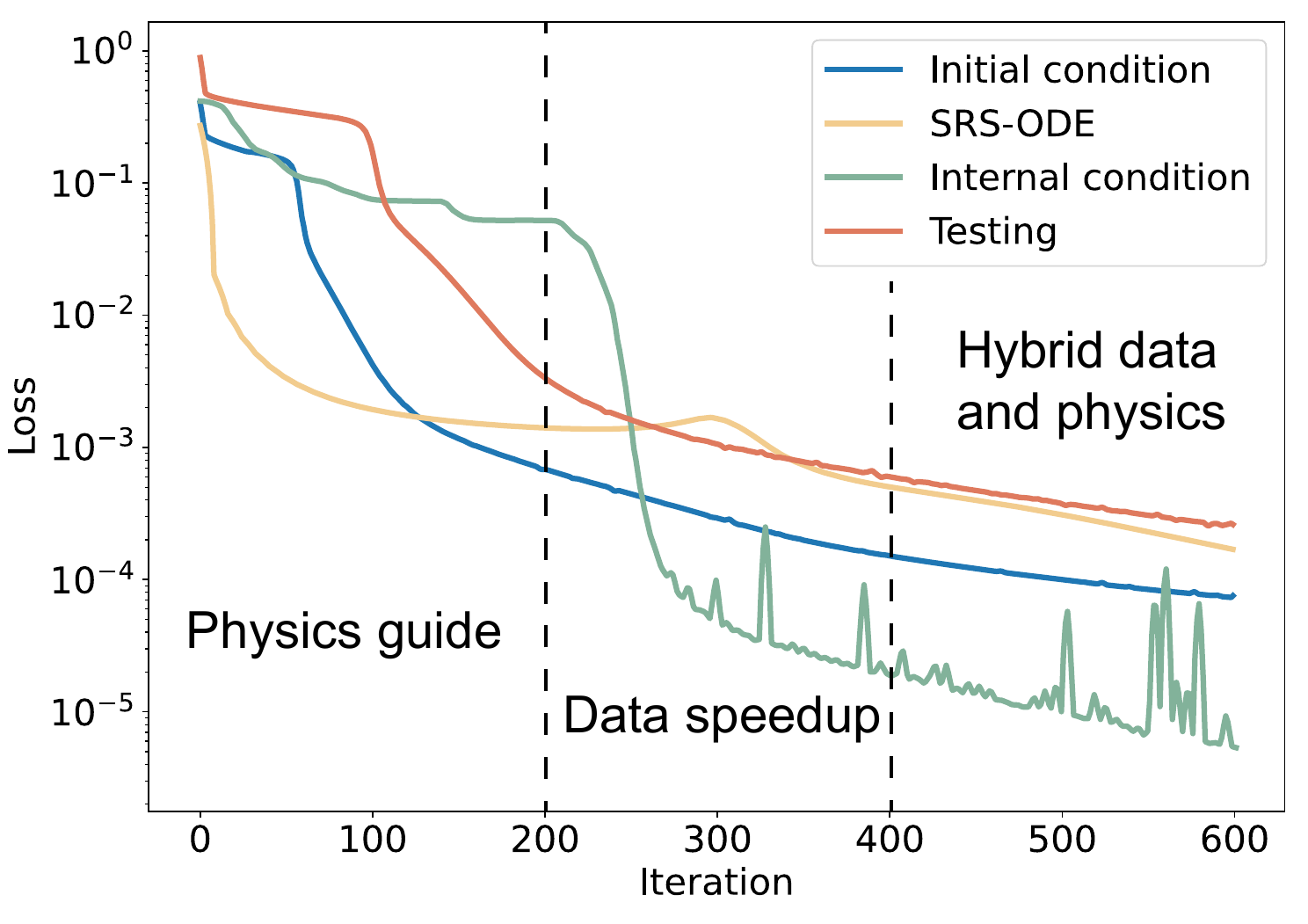}
\caption{Training loss of PEO in the forward power prediction.}
\label{fig6}
\end{figure}
\begin{figure}[t]
\centering
\includegraphics[width=0.48\textwidth]{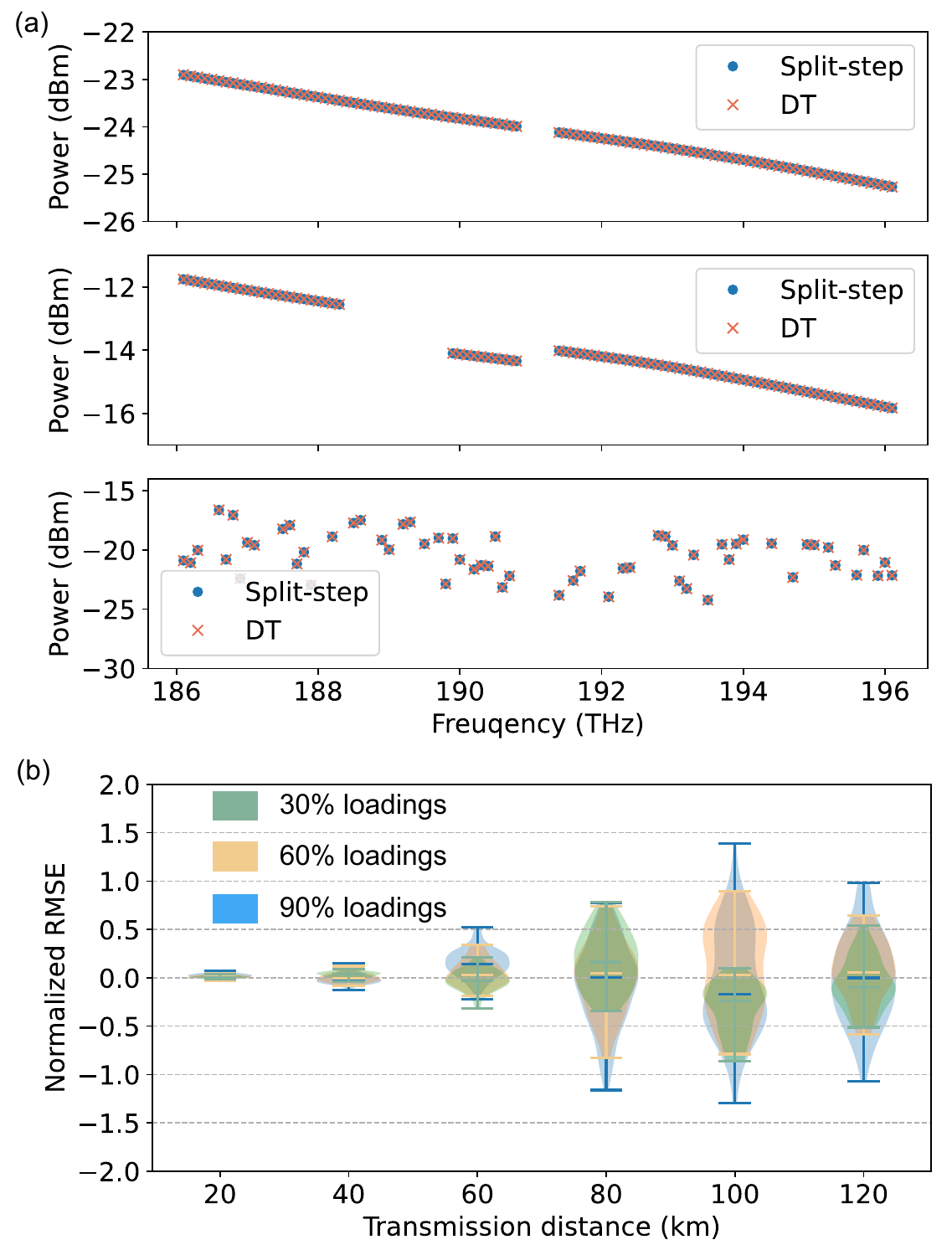}
\caption{Simulation results of (a) power profiles of three examples of forward prediction and (b) normalized RMSE of tests on different loadings and transmission distances.}
\label{fig7}
\end{figure}

The amplifier determines the accumulation of ASE noise power $P_{ASE}$. As the next-generation optical networks cover wider transmission bandwidth, an adequate handling of the variations along the frequency of the network physical features becomes crucial. For EDFA modeling, frequency-dependent profiles of gain and NF are used for optical signal amplification \cite{song2022efficient}. Noise induced by transceivers is another noticeable impairment in short-reach systems. The $SNR_{TRX}$ is derived from the back-to-back implementation penalty observed in experiments, as indicated in \cite{buglia2022impact}. Additionally, WSSs integrated in the ROADMs can induce penalties on the transmitted optical signal due to the imperfect filtering. In this paper, the filtering penalty is accounted for by the penalty coefficient $a$, which is approximated by the difference between the predicted SNR results (including impairments of ASE noise, NLI, and noise from transceivers) and measured ones from the actual link situation \cite{RN2, buglia2022impact}.

\subsection{Evaluation of DT in forward power prediction}
For the training dataset of DT, we randomly generate 6,000 input power profiles on C+L-band with the number of channels ranging from 10 to 96 and the power of each channel ranging from -6 to 9 dBm. All these initial conditions are used for physics-informed training, and for the other part of data-driven training, we collect corresponding outputs of 10\% of these initial conditions with transmission distance at 20/40/60/80 and 100 km. Following the three training stages outlined in Fig. \ref{fig4}(a), the losses for the hybrid data-driven and physics-informed training process of this DT are shown in Fig. \ref{fig6}. During the first step of physics guidance, we observe a rapid decrease in the SRS-ODE loss. However, the internal loss with labels at 20/40/60/80 and 100 km does not decrease 
correspondingly, indicating differing objectives between physics-informed and data-driven training. This discrepancy is similarly observed during step 2, where data-driven speedup is applied, yet the SRS-ODE loss even experiences a slight increase. Ultimately, during the hybrid data and physics training stage, all losses converge to a low point, signifying successful integration of both approaches. It should be noted that it is impossible to train this DT in a pure data-driven manner using a small data set with only 600 pairs of labeled data. The integration of physics-informed training greatly enhances the generalization ability.

Approximately 1 hour was paid in the training process with GPU Tesla T4. The number of neural nodes in each layer is [96 200 200 200 96] for the BraNet and [1 100 100 96] for the TruNet. However, once trained, the DT can generalize well to unseen input power profiles thanks to the guidance of physical laws. As depicted in Fig. \ref{fig5}, we illustrate one link in the COST239 network as an example. In this specific link, spanning from node 8 to node 11, the first span has a fiber length of 100km, while the subsequent 9 spans have a fiber length of 80km each. Homogenous spans are used for other links in this network. For each span, the separate EDFA on C and L band compensate for the fiber loss and connector loss. Fig. \ref{fig7}(a) reports testing results after 120km transmission, with one full loading of uniform 1mW launch power, a case of 20 continuous channels cut on the L-band and a case of 80\% random loading. All results agree well with the numerical split-step methods of 100m step size, and the statistical test results are displayed in Fig. \ref{fig7}(b). For testing cases with 1,000 new input power profiles with random loadings, the normalized root mean-square-error (RMSE) generally falls in 1x$10^{-4}$ within 120km. It can be observed that the accuracy decreases a bit with longer distance and more channels, where the impairments on the power profiles can be more complex to be learned by the DT. 

\begin{figure}[t]
\centering
\includegraphics[width=0.47\textwidth]{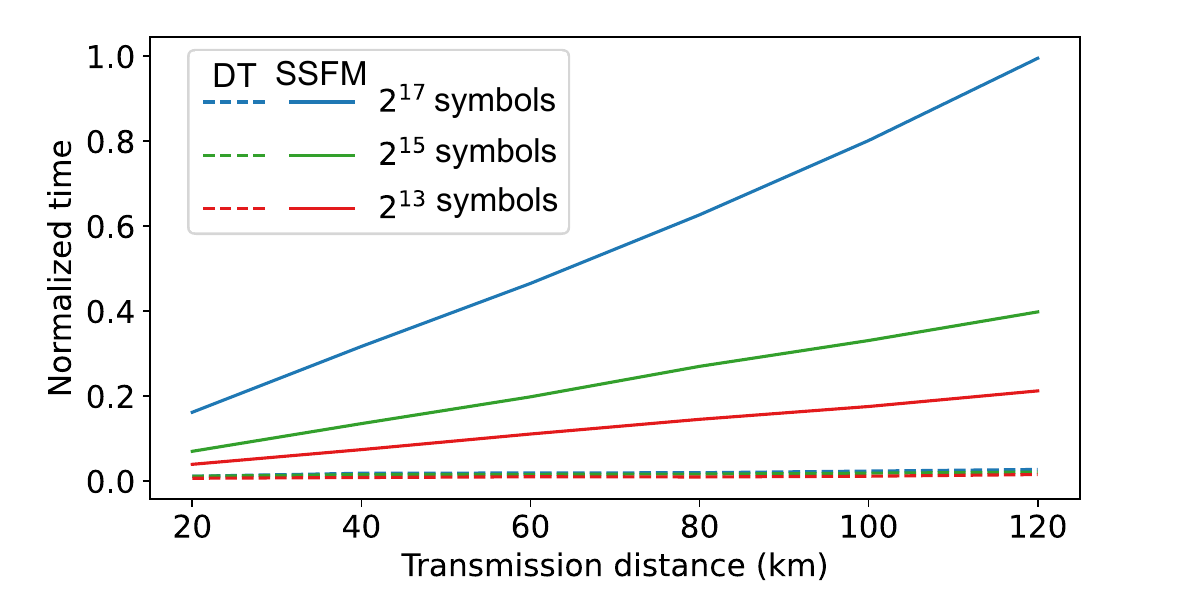}
\caption{Normalized time for performance estimation with different number of symbols and transmission distances. The time is normalized by the time cost of SSFM at 120km with $2^{17}$ symbols.}
\label{fig8}
\end{figure}

The established DT exhibits faster calculation speed compared to conventional split-step method with iterative split-step calculation. The calculation time required by split-step method increases roughly linearly with the number of steps, which in turn increases with the transmission distance. However, the PEO provides a closed-form solution for forward power prediction, eliminating the need for iterative steps, and its calculation time does not increase with transmission distance within a span. The calculation time for transmission distances within a 120 km span and different volumes of transmitted symbols is illustrated in Fig. \ref{fig8}. It can be observed that the calculation time can be reduced by up to 100 times using the closed-form DT compared to split-step method at 120 km with $2^{17}$ symbols, where the step size of split-step method is 100m and requires iterative calculation of 1,200 steps. For the entire simulation network, split-step method requires approximately 160 seconds, while the DT requires only around 1.5 seconds under the same computational environment. With the hybrid training approach, the established DT serves as a fast yet accurate solver for channel performance prediction. 


\begin{figure*}[t]
\centering\includegraphics[width=1\textwidth]{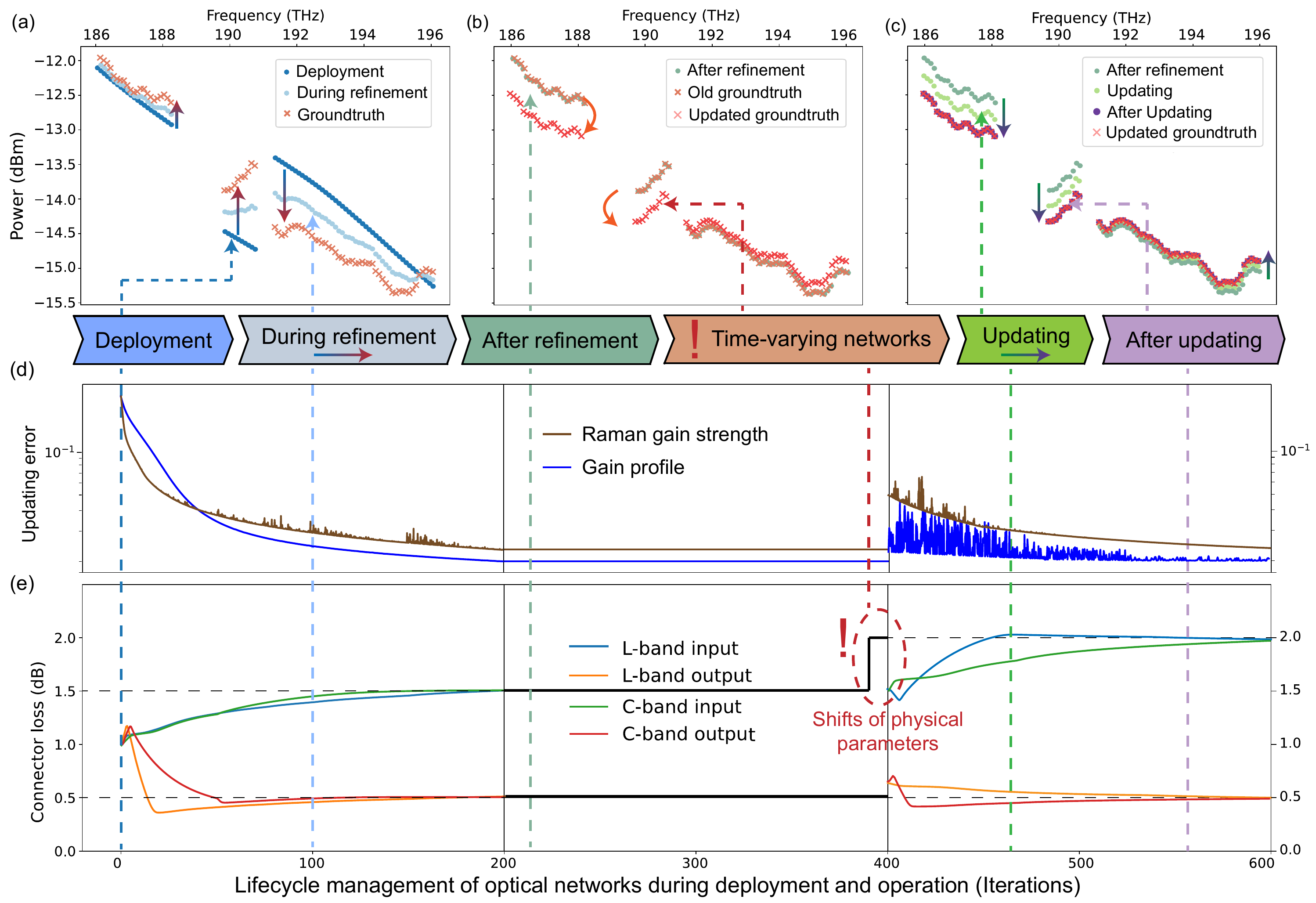}
\caption{Results of using dynamic-updating DT on the lifecycle management of optical networks. (a-c) Channel power profiles on different stages. (d) Updating error of Raman gain strength and EDFA gain profile. (d) Convergence of connector loss.}
\label{fig10}
\end{figure*}

\subsection{Lifecycle management of optical networks using dynamic-updating DT}
To showcase the application of dynamic-updating DT in the lifecycle management of optical networks, we conduct parameter refinement during the installation and deployment stage and implement parameter updating following network changes in the afterwards operation and maintenance stage. Once trained offline, the DT can be employed for forward prediction without requiring further training. At the deployment stage, physical parameters are refined with the measured input and output power profiles at $z=0$ and $z_{max}$, respectively. Then, the output optical performance can be predicted using the updated DT, which can be used for optimization, upgrades, or fault management of field-deployed networks. At the operation and maintenance stage, the predicted output profiles are compared with the measured ones regularly at critical points within the optical networks. When the power error of any channel exceeds the threshold (for example, 0.5dB), the measured power profiles at the corresponding point should be collected by OCMs, and the DT updating cycle is activated. With the measured input and output power profiles of the corresponding span, physical parameters $\Lambda$ can be updated, as discussed in Section IV. In field-deployed optical networks, channel power profiles can be easily collected by OCMs. This iterative process ensures the continuous refinement and updating of the DT to adapt to evolving network conditions throughout its operational lifetime.

The dynamic-updating DT is initially trained with default fiber parameters, wherein the normalized Raman gain strength is set to 1 for each fiber, the connector loss is 1 dB for C and L band at each connection, and a linear gain profile is assumed for each EDFA. We continue to take the 10-span link between node 8 and 11 as an example. For the first span of this link, the actual connector loss is set to be 1.5dB at the input of fiber and 0.5dB at the output of fiber for both C and L-band. The actual Raman gain strength is 1.2 and the gain profile for C and L-band EDFA is frequency-dependent. Upon deployment of this offline-trained dynamic-updating DT at the greenfield stage of optical networks, parameter refinement should be conducted to transition from a type-specialized DT, which has default physical parameters for devices of the same type, to an instance-specialized DT, which possesses refined physical parameters for each instance. As shown in Fig. \ref{fig10}, the first 200 iterations are dedicated to parameter refinement upon deployment of the DT. During this process, the input connector loss of both C- and L-band converges from 1 to the actual value of 1.5dB, while the output connector loss converges from 1 to 0.5dB.

Simultaneously, the EDFA gain profiles are refined from linear to frequency-dependent ones. The error of the refined Raman gain strength and gain profile during the first 200 iterations is depicted in Fig. \ref{fig10}(d). As depicted in Fig. \ref{fig10}(a), at the start point, the power profiles predicted by the dynamic-updating DT do not match the ground truth values. However, through parameter refinement, the DT is able to produce accurate results, as demonstrated in Fig. \ref{fig10}(b). Once the optical networks are stable, the DT will operate appropriately, and the predictions align closely with the actual measurements. In the simulations, we assume the shift of some fiber parameters due to ageing or unexpected human activities. As the environments changing during the operation, these physical parameters can also shift. When the discrepancy between the channel powers predicted by the PEO and the measured ones exceeds 0.5 dB for any channel, the dynamic-updating DT initiates dynamic updating. In this scenario of time-varying networks, we introduce a change where the input connector loss for both C- and L-bands shifts from 1.5 dB to 2 dB, as depicted in Fig. \ref{fig10}(b). Subsequently, the DT initiates dynamic updating, leading to the convergence of the connector loss of the fiber input to 2 dB. However, other physical parameters may experience temporary deviations before returning to their original correct values, as depicted by some ripples in the updates. With only a few pairs of measured channel power profiles collected by OCMs, the real-time values of these parameters can be updated with minimal costs, ensuring the continued accuracy and reliability of the DT in adapting to changing network conditions.

\section{Demonstration on field-deployed optical networks}
\begin{figure*}[t]
\centering\includegraphics[width=18cm]{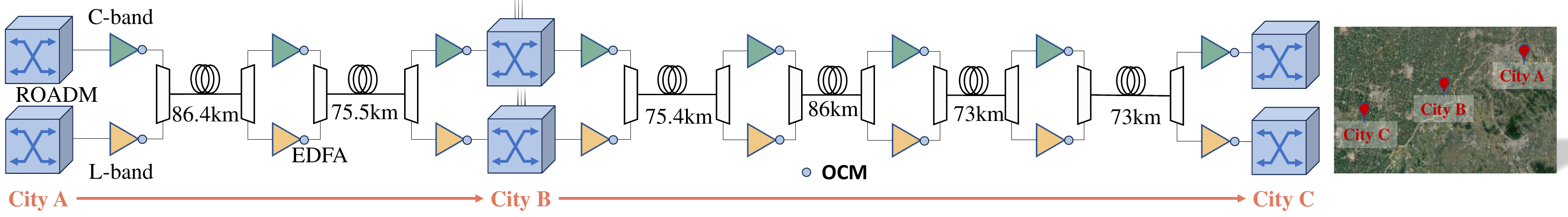}
\caption{Schematic of field-deployed C+L-band transmission link connecting three cities, and schematic of geographical location of these cities.}
\label{fig11}
\end{figure*}

To demonstrate the capability of our proposed dynamic-updating DT in real environments, where various practical restrictions and uncertainties exist, we further tested it on a field-deployed link. The field-trial C48+L48 WDM transmission link consists of six amplified spans with a maximum length of 86.4km (totaling 469.3km of G.652 SMF) as shown in Fig. \ref{fig11}. A ROADM station is placed at the third site in city B, while in other sites, in-line EDFA is used for separated amplification of C- and L-band. OCM is placed at the end of each EDFA for channel power collection. Three commercial 400Gb/s transponders on the C-band and two on the L-band are configured for five channels under test (CUT), and probabilistic constellation shaping-quadrature amplitude modulation (PCS-16QAM) with 91.6 baud rate is modulated for optical transmission with 100GHz channel spacing. In the transceiver side, signals are MUX/DEMUX by ROADM, and other channels are filled with filtered ASE noise for full load configuration on C+L-band at start. These ASE channels can be deactivated at will to configure partial loading. The transmission bandwidth occupies the L-band, from 186.1 THz to 190.8 THz, and the C-band, from 191.4 THz to 196.1 THz with a total of 96 channels. In addition, the NF of the deployed commercial EDFAs are estimated from measured OSNR values, and the impairments induced by the five commercial transceivers of CUT are measured under BtB setup \cite{RN2} in advance. In deployed systems, data such as routing and spectrum information, topology, device configuration, and physical parameters are collected from the NOS using telemetry techniques \cite{paolucci2018network}. These data are then merged into a standard format in NOS for further processing and analysis \cite{lara2013network}. For the final optical performance derived by the DT, polarization-dependent loss (PDL) introduced by ROADMs and EDFA, and the polarization mode dispersion (PMD) introduced by the fiber channel are considered into the penalty coefficient in Eq. (\ref{eq1}). With data of channel power profiles collected from NOS, the learning of proposed DT can be conducted. It should be noted that the proposed dynamic-updating DT operates offline, separate from the NOS, with learning conducted on separate computers distinct from those controlling the NOS.

\begin{figure*}[t]
\centering\includegraphics[width=18cm]{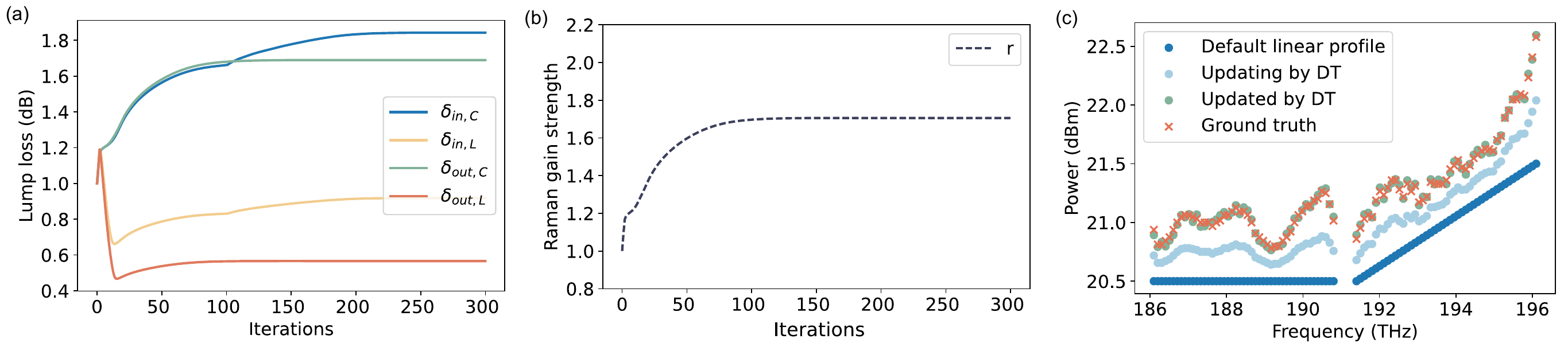}
\caption{Parameter updating trace of (a) connector loss, (b) Raman gain strength, and (c) gain profiles in the field-trial transmission link.}
\label{fig12}
\end{figure*}

\begin{figure*}[t]
\centering\includegraphics[width=18cm]{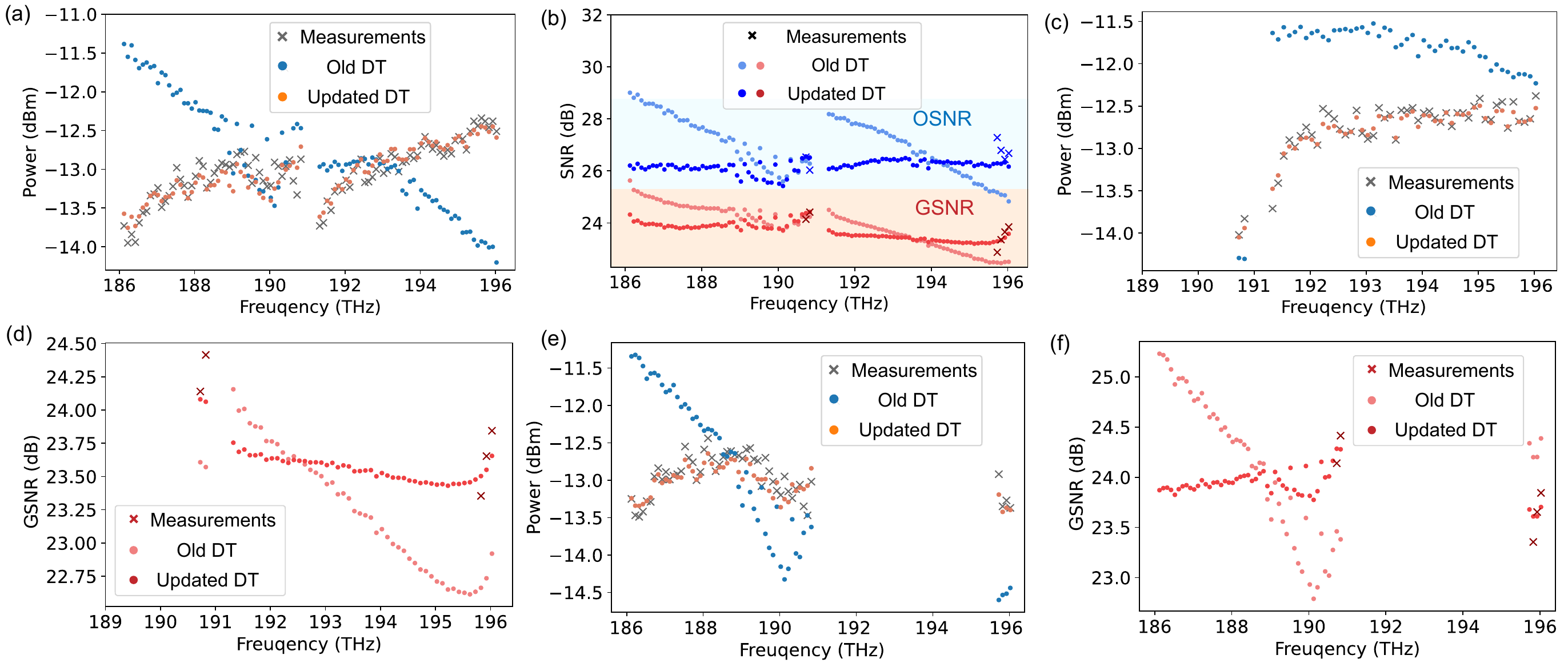}
\caption{Predictions of (a) channel power and (b) OSNR and GSNR on full loading condition, (c) channel power and (d) GSNR on L-band partial loading condition, and (e) channel power and (f) GSNR on C-band partial loading condition. Measurements refer to the channel powers measured by the OCM or SNR derived from measured BER.}
\label{fig13}
\end{figure*}

This transmission link has been operational for some time, leading to changes in the EDFA characteristics. During the brownfield operation and maintenance phase, we conduct a hypothetical EDFA replacement due to ageing, multiple re-pluggings, and other factors. This replacement alters the values of connection losses and amplifier gain profiles. The changed values can be measured using additional devices in this first span, such as power meters and OCM. Additionally, the fiber in the first span is replaced as part of a protection switching operation, as the original fiber path was cut. Such scenarios are typical tasks in the lifecycle management of optical networks. To update the DT after device replacement, we focus on the first span of this operational field-trial link to illustrate the dynamic updating process during the brownfield stage. The coarse data-sheet parameters of $\Lambda$ for the first span are $\delta_{in(out),C(L)}$=1dB, $r$=1, and point-of-manufacturer type-specialized EDFA linear gain profile with correction bias. Eight pairs of channel powers before and after this span measured by OCM along the regular operations are used as initial and final conditions.

The updating trace of these refined parameters are depicted in Fig. \ref{fig12}. This process can be done parallelly for each span. It can be observed that the connection loss is updated from 1dB to around 1.8 and 0.9dB for the input of C-and L-band, and to around 1.7 and 0.5dB for the output of C- and L-band, respectively. The Raman gain strength is updated from 1 to around 1.7, which means the actual value of Raman gain is larger than the recorded ones on handbook. For the EDFA, the gain tilt on the C-band is 1dB. Under this situation, the gain profile is updated from linear ones to frequency-dependent ones as shown in Fig. \ref{fig12}(c). With these refined parameters, the prediction accuracy for channel power and QoT can be improved. The results for full loading are shown in Fig. \ref{fig13}(a) and (b), and the average accuracy of power prediction is improved from 1.1 to 0.12 (RMSE in dB units) with a per-channel accuracy improvement of 0.8dB in average and 2.4dB in max. For CUT, the measured OSNR are delivered from controller, and the GSNR is derived from pre-FEC BER. The maximum per channel accuracy improvement of OSNR and GSNR for CUT is 1.6 and 1.4dB, respectively. For partial loading of C- and L-band, where the ASE channels are removed, the accuracy is overall improved as shown in Fig. \ref{fig13}(c)-(d). Performance improvement on different loadings imply that the proposed dynamic-updating DT is able to facilitate the management and control of field-deployed optical networks. Moreover, inaccurate measurements present significant challenges, particularly in field-deployed optical networks. Measurements with error bias can be corrected if the bias is identified through multiple tests or comparisons. However, measurement jitter, which arises from the limitations of measurement devices or environmental conditions, is difficult to predict and compensate for, making it a more intricate challenge to address.

\section{Conclusion}
In this paper, the dynamic-updating DT trained by hybrid data-driven and physics-informed approach is demonstrated for the lifecycle management of optical networks from the installation and deployment stage to the operation and maintenance stage. For DT, we focus on the dynamic updating feature, which is often overlooked in simulations and experiments with near-ideal environments. In a field-trial C+L-band transmission link, the dynamic-updating DT is demonstrated for physical parameter updating after device replacement. Improved prediction accuracy is observed for channel power and performance. Although we can only monitor data from this field-deployed link for a limited period of time, the lifecycle management using DT from deployment to the operation stage is illustrated through simulations. There are still some open challenges on the implementation of dynamic-updating DT on field-deployed optical networks. First, in addition to identifying and updating frequency-dependent physical parameters, it is crucial to consider longitudinal parameters in the proposed DT, particularly for detecting anomalies along the link. Furthermore, transceiver characteristics, filtering penalties, and EDFA noise levels play significant roles and should also be updated for more accurate prediction. Moving forward, a critical next step for the DT is its integration into the SDN controller to demonstrate a more practical application for the full lifecycle management of real-world networks. The learning is expected to be performed within the NOS, as the learning process is not data- or computation-intensive. The dynamic-updating DT is expected to drive the next-generation optical networks towards lifecycle autonomous management. Additionally, the proposed hybrid approach, combining data-driven techniques with physics-informed neural operator networks, holds promise for applications beyond optical networks, particularly in areas requiring self-updating capabilities.

\section*{Acknowledgments}
The authors would like to acknowledge the China Unicom for providing the field-trial testbed for this research.


\bibliography{DUDT}

\begin{thebibliography}{10}
\providecommand{\url}[1]{#1}
\csname url@samestyle\endcsname
\providecommand{\newblock}{\relax}
\providecommand{\bibinfo}[2]{#2}
\providecommand{\BIBentrySTDinterwordspacing}{\spaceskip=0pt\relax}
\providecommand{\BIBentryALTinterwordstretchfactor}{4}
\providecommand{\BIBentryALTinterwordspacing}{\spaceskip=\fontdimen2\font plus
\BIBentryALTinterwordstretchfactor\fontdimen3\font minus \fontdimen4\font\relax}
\providecommand{\BIBforeignlanguage}[2]{{%
\expandafter\ifx\csname l@#1\endcsname\relax
\typeout{** WARNING: IEEEtran.bst: No hyphenation pattern has been}%
\typeout{** loaded for the language `#1'. Using the pattern for}%
\typeout{** the default language instead.}%
\else
\language=\csname l@#1\endcsname
\fi
#2}}
\providecommand{\BIBdecl}{\relax}
\BIBdecl

\bibitem{RN155}
Y.~Pointurier, ``Design of low-margin optical networks,'' \emph{Journal of Optical Communications and Networking}, vol.~9, no.~1, pp. A9--A17, 2017.

\bibitem{mitra2019effect}
A.~Mitra, D.~Semrau, N.~Gahlawat, A.~Srivastava, P.~Bayvel, and A.~Lord, ``Effect of reduced link margins on c+ l band elastic optical networks,'' \emph{Journal of Optical Communications and Networking}, vol.~11, no.~10, pp. C86--C93, 2019.

\bibitem{essiambre2010capacity}
R.-J. Essiambre, G.~Kramer, P.~J. Winzer, G.~J. Foschini, and B.~Goebel, ``Capacity limits of optical fiber networks,'' \emph{Journal of Lightwave technology}, vol.~28, no.~4, pp. 662--701, 2010.

\bibitem{hoshida2022ultrawideband}
T.~Hoshida, V.~Curri, L.~Galdino, D.~T. Neilson, W.~Forysiak, J.~K. Fischer, T.~Kato, and P.~Poggiolini, ``Ultrawideband systems and networks: Beyond c+ l-band,'' \emph{Proceedings of the IEEE}, vol. 110, no.~11, pp. 1725--1741, 2022.

\bibitem{stepanovsky2019comparative}
M.~Stepanovsky, ``A comparative review of mems-based optical cross-connects for all-optical networks from the past to the present day,'' \emph{IEEE Communications Surveys \& Tutorials}, vol.~21, no.~3, pp. 2928--2946, 2019.

\bibitem{soumplis2017network}
P.~Soumplis, K.~Christodoulopoulos, M.~Quagliotti, A.~Pagano, and E.~Varvarigos, ``Network planning with actual margins,'' \emph{Journal of Lightwave Technology}, vol.~35, no.~23, pp. 5105--5120, 2017.

\bibitem{RN158}
E.~J. Tuegel, A.~R. Ingraffea, T.~G. Eason, and S.~M. Spottswood, ``Reengineering aircraft structural life prediction using a digital twin,'' \emph{International Journal of Aerospace Engineering}, vol. 2011, 2011.

\bibitem{RN120}
L.~Wright and S.~Davidson, ``How to tell the difference between a model and a digital twin,'' \emph{Advanced Modeling and Simulation in Engineering Sciences}, vol.~7, no.~1, 2020.

\bibitem{RN122}
M.~S. Faruk and S.~J. Savory, ``Measurement informed models and digital twins for optical fiber communication systems,'' \emph{Journal of Lightwave Technology}, vol.~42, no.~3, pp. 1016--1030, 2024.

\bibitem{wang2021role}
D.~Wang, Z.~Zhang, M.~Zhang, M.~Fu, J.~Li, S.~Cai, C.~Zhang, and X.~Chen, ``The role of digital twin in optical communication: fault management, hardware configuration, and transmission simulation,'' \emph{IEEE Communications Magazine}, vol.~59, no.~1, pp. 133--139, 2021.

\bibitem{vilalta2023applying}
R.~Vilalta, L.~Gifre, R.~Casellas, R.~Mu{\~n}oz, R.~Mart{\'\i}nez, A.~Mozo, A.~Pastor, D.~L{\'o}pez, and J.~P. Fern{\'a}ndez-Palacios, ``Applying digital twins to optical networks with cloud-native sdn controllers,'' \emph{IEEE Communications Magazine}, 2023.

\bibitem{RN2}
V.~Curri, ``Digital-twin of physical-layer as enabler for open and disaggregated optical networks,'' in \emph{2023 International Conference on Optical Network Design and Modeling (ONDM)}.\hskip 1em plus 0.5em minus 0.4em\relax IEEE, Conference Proceedings, pp. 1--6.

\bibitem{RN61}
C.~Janz, Y.~You, M.~Hemmati, Z.~Jiang, A.~Javadtalab, and J.~Mitra, ``Digital twin for the optical network: Key technologies and enabled automation applications,'' in \emph{NOMS 2022-2022 IEEE/IFIP Network Operations and Management Symposium}.\hskip 1em plus 0.5em minus 0.4em\relax IEEE, Conference Proceedings, pp. 1--6.

\bibitem{RN156}
Y.~Song, M.~Zhang, Y.~Zhang, Y.~Shi, S.~Shen, B.~Guo, S.~Huang, and D.~Wang, ``Implementing digital twin in field-deployed optical networks: Uncertain factors, operational guidance, and field-trial demonstration,'' \emph{IEEE Network}, 2023.

\bibitem{RN157}
A.~Ferrari, K.~Balasubramanian, M.~Filer, Y.~Yin, E.~Le~Rouzic, J.~Kundrát, G.~Grammel, G.~Galimberti, and V.~Curri, ``Assessment on the in-field lightpath qot computation including connector loss uncertainties,'' \emph{Journal of Optical Communications and Networking}, vol.~13, no.~2, p. A156, 2020.

\bibitem{pesic2019impact}
J.~Pesic, N.~Rossi, and T.~Zami, ``Impact of margins evolution along ageing in elastic optical networks,'' \emph{Journal of Lightwave Technology}, vol.~37, no.~16, pp. 4081--4089, 2019.

\bibitem{song2023physics}
Y.~Song, M.~Zhang, Y.~Shi, Y.~Tang, Y.~Hu, S.~Shen, and D.~Wang, ``Physics-informed digital twin with parameter refinement for a field-trial c+ l-band transmission link,'' in \emph{49th European Conference on Optical Communications (ECOC 2023)}, vol. 2023.\hskip 1em plus 0.5em minus 0.4em\relax IET, 2023, pp. 1190--1193.

\bibitem{wang2024digital}
D.~Wang, Y.~Song, Y.~Zhang, X.~Jiang, J.~Dong, F.~N. Khan, T.~Sasai, S.~Huang, A.~P.~T. Lau, M.~Tornatore \emph{et~al.}, ``Digital twin of optical networks: A review of recent advances and future trends,'' \emph{Journal of Lightwave Technology}, 2024.

\bibitem{mackay2022field}
A.~W. MacKay and D.~W. Boertjes, ``Field learnings of deploying model assisted network feedback systems,'' in \emph{Optical Fiber Communication Conference}.\hskip 1em plus 0.5em minus 0.4em\relax Optica Publishing Group, 2022, pp. W4G--2.

\bibitem{morette2023machine}
N.~Morette, H.~Hafermann, Y.~Frignac, and Y.~Pointurier, ``Machine learning enhancement of a digital twin for wavelength division multiplexing network performance prediction leveraging quality of transmission parameter refinement,'' \emph{Journal of Optical Communications and Networking}, vol.~15, no.~6, pp. 333--343, 2023.

\bibitem{RN108}
J.~Zhou, J.~Lu, and C.~Yu, ``Improving the accuracy of qot estimation with insertion loss distribution evaluation for c + l band transmission systems,'' \emph{Journal of Optical Communications and Networking}, vol.~16, no.~1, p.~12, 2023.

\bibitem{RN109}
Y.~He, Z.~Zhai, L.~Dou, L.~Wang, Y.~Yan, C.~Xie, C.~Lu, and A.~P.~T. Lau, ``Improved qot estimations through refined signal power measurements and data-driven parameter optimizations in a disaggregated and partially loaded live production network,'' \emph{Journal of Optical Communications and Networking}, vol.~15, no.~9, p. 638, 2023.

\bibitem{RN160}
O.~Karandin, A.~Ferrari, F.~Musumeci, Y.~Pointurier, and M.~Tornatore, ``Probabilistic low-margin optical-network design with multiple physical-layer parameter uncertainties,'' \emph{Journal of Optical Communications and Networking}, vol.~15, no.~7, p. C129, 2023.

\bibitem{RN161}
T.~Sasai, E.~Yamazaki, and Y.~Kisaka, ``Performance limit of fiber-longitudinal power profile estimation methods,'' \emph{Journal of Lightwave Technology}, 2023.

\bibitem{RN162}
G.~E. Karniadakis, I.~G. Kevrekidis, L.~Lu, P.~Perdikaris, S.~Wang, and L.~Yang, ``Physics-informed machine learning,'' \emph{Nature Reviews Physics}, 2021.

\bibitem{RN163}
M.~Raissi, A.~Yazdani, and G.~E. Karniadakis, ``Hidden fluid mechanics: Learning velocity and pressure fields from flow visualizations,'' \emph{Science}, vol. 367, no. 6481, pp. 1026--1030, 2020.

\bibitem{RN164}
X.~Jiang, D.~Wang, Q.~Fan, M.~Zhang, C.~Lu, and A.~P.~T. Lau, ``Physics‐informed neural network for nonlinear dynamics in fiber optics,'' \emph{Laser Photonics Reviews}, vol.~16, no.~9, p. 2100483, 2022.

\bibitem{song2022physics}
Y.~Song, D.~Wang, Q.~Fan, X.~Jiang, X.~Luo, and M.~Zhang, ``Physics-informed neural operator for fast and scalable optical fiber channel modelling in multi-span transmission,'' in \emph{2022 European Conference on Optical Communication (ECOC)}.\hskip 1em plus 0.5em minus 0.4em\relax IEEE, 2022, pp. 1--4.

\bibitem{lu2021physics}
L.~Lu, R.~Pestourie, W.~Yao, Z.~Wang, F.~Verdugo, and S.~G. Johnson, ``Physics-informed neural networks with hard constraints for inverse design,'' \emph{SIAM Journal on Scientific Computing}, vol.~43, no.~6, pp. B1105--B1132, 2021.

\bibitem{RN28}
Y.~Song, Y.~Zhang, C.~Zhang, J.~Li, M.~Zhang, and D.~Wang, ``Pinn for power evolution prediction and raman gain spectrum identification in c+ l-band transmission system,'' in \emph{2023 Optical Fiber Communications Conference and Exhibition (OFC)}.\hskip 1em plus 0.5em minus 0.4em\relax IEEE, Conference Proceedings, pp. 1--3.

\bibitem{song2024srs}
Y.~Song, M.~Zhang, X.~Jiang, F.~Zhang, C.~Ju, S.~Huang, A.~P.~T. Lau, and D.~Wang, ``Srs-net: a universal framework for solving stimulated raman scattering in nonlinear fiber-optic systems by physics-informed deep learning,'' \emph{Communications Engineering}, vol.~3, no.~1, p. 109, 2024.

\bibitem{wang2020data}
D.~Wang, Y.~Song, J.~Li, J.~Qin, T.~Yang, M.~Zhang, X.~Chen, and A.~C. Boucouvalas, ``Data-driven optical fiber channel modeling: A deep learning approach,'' \emph{Journal of Lightwave Technology}, vol.~38, no.~17, pp. 4730--4743, 2020.

\bibitem{lu2021learning}
L.~Lu, P.~Jin, G.~Pang, Z.~Zhang, and G.~E. Karniadakis, ``Learning nonlinear operators via deeponet based on the universal approximation theorem of operators,'' \emph{Nature machine intelligence}, vol.~3, no.~3, pp. 218--229, 2021.

\bibitem{wang2021learning}
S.~Wang, H.~Wang, and P.~Perdikaris, ``Learning the solution operator of parametric partial differential equations with physics-informed deeponets,'' \emph{Science advances}, vol.~7, no.~40, p. eabi8605, 2021.

\bibitem{buglia2022impact}
H.~Buglia, E.~Sillekens, A.~Vasylchenkova, P.~Bayvel, and L.~Galdino, ``On the impact of launch power optimization and transceiver noise on the performance of ultra-wideband transmission systems,'' \emph{Journal of Optical Communications and Networking}, vol.~14, no.~5, pp. B11--B21, 2022.

\bibitem{borraccini2023experimental}
G.~Borraccini, S.~Straullu, A.~Giorgetti, R.~Ambrosone, E.~Virgillito, A.~D’Amico, R.~D’Ingillo, F.~Aquilino, A.~Nespola, N.~Sambo \emph{et~al.}, ``Experimental demonstration of partially disaggregated optical network control using the physical layer digital twin,'' \emph{IEEE Transactions on Network and Service Management}, 2023.

\bibitem{song2022efficient}
Y.~Song, Q.~Fan, C.~Lu, D.~Wang, and A.~P.~T. Lau, ``Efficient three-step amplifier configuration algorithm for dynamic c+ l-band links in presence of stimulated raman scattering,'' \emph{Journal of Lightwave Technology}, vol.~41, no.~5, pp. 1445--1453, 2022.

\bibitem{paolucci2018network}
F.~Paolucci, A.~Sgambelluri, F.~Cugini, and P.~Castoldi, ``Network telemetry streaming services in sdn-based disaggregated optical networks,'' \emph{Journal of Lightwave Technology}, vol.~36, no.~15, pp. 3142--3149, 2018.

\bibitem{lara2013network}
A.~Lara, A.~Kolasani, and B.~Ramamurthy, ``Network innovation using openflow: A survey,'' \emph{IEEE communications surveys \& tutorials}, vol.~16, no.~1, pp. 493--512, 2013.

\end{thebibliography}
\bibliographystyle{IEEEtran}

\vfill

\end{document}